\documentclass[journal]{IEEEtran}

\ifCLASSINFOpdf
\else
   \usepackage[dvips]{graphicx}
\fi
\usepackage{url}

\usepackage{graphicx}
\usepackage{amsmath}
\usepackage{amssymb}
\usepackage{booktabs}
\usepackage{multirow}
\usepackage{pifont}
\usepackage{wrapfig}
\usepackage{cite}
\usepackage{color, colortbl}
	
\definecolor{Gray}{gray}{0.9}

\begin{document}
\newcommand{\E}{\mathbb{E}}
\newcommand{\R}{\mathbb{R}}
\newcommand{\calD}{\mathcal{D}}
\newcommand{\calL}{\mathcal{L}}
\newcommand{\calT}{\mathcal{T}}
\newcommand{\calX}{\mathcal{X}}
\newcommand{\calY}{\mathcal{Y}}
\newcommand{\bx}{\mathbf{x}}
\newcommand{\by}{\mathbf{y}}
\newcommand{\bz}{\mathbf{z}}
\newcommand{\bC}{\mathbf{C}}
\newcommand{\bZ}{\mathbf{Z}}
\newcommand{\argmin}{\operatorname*{argmin}}
\newcommand{\argmax}{\operatorname*{argmax}}
\newcommand{\similarity}{\text{sim}}
\newcommand{\cross}{\ding{55}}

\title{Learning Representations for New Sound Classes With Continual Self-Supervised Learning}

\author{Zhepei Wang$^\sharp$, Cem Subakan$^{\natural \flat}$, Xilin Jiang$^\dagger$, Junkai Wu$^\sharp$, Efthymios Tzinis$^\sharp$, Mirco Ravanelli$^{\natural \chi \flat}$, \\ and Paris Smaragdis$^{\sharp \ddag}$, \IEEEmembership{Fellow, IEEE}
\vspace{-.5cm} 

\thanks{This manuscript is submitted for review on 10/03/2022. This work was supported in part by NIFA award \#2020-67021-32799. }

\thanks{$\sharp$ University of Illinois at Urbana-Champaign, USA;  $\natural$ Concordia University, Canada;  $\chi$ Université de Montréal, Canada.; $\dagger$ Columbia University; $\flat$ Mila-Quebec AI Institute, Canada; $\ddag$ Amazon Web Services, USA.}

}

\markboth{}
{Shell \MakeLowercase{\textit{et al.}}: Bare Demo of IEEEtran.cls for IEEE Journals}
\maketitle

\newcommand{\cem}[1]{#1}
\newcommand{\cemCR}[1]{#1}
\newcommand{\zp}[1]{#1}
\newcommand{\zpCR}[1]{#1}

\begin{abstract}






\cem{ 
In this paper, we work on a sound recognition system that continually incorporates new sound classes. Our main goal is to develop a framework where the model can be updated without relying on labeled data. For this purpose, we propose adopting representation learning, where an encoder is trained using unlabeled data. This learning framework enables the study and implementation of a practically relevant use case where only a small amount of the labels is available in a continual learning context. We also make the empirical observation that a similarity-based representation learning method within this framework is robust to forgetting even if no explicit mechanism against forgetting is employed. We show that this approach obtains similar performance compared to several distillation-based continual learning methods when employed on self-supervised representation learning methods.}




\end{abstract}


\IEEEpeerreviewmaketitle

\section{Introduction}
\label{sec:intro}

\IEEEPARstart{D}{eep} neural networks for audio classification require large amounts of data to be trained on \cite{pann, ssast}. However, in many realistic deployments, additional labeled data for new sound classes might present itself after initial training, necessitating a time and resource-consuming retraining process that incorporates both past and new data. This problem could be exacerbated by storage constraints, hardware corruption, or privacy regulations that can limit access to past data. 



Continual/lifelong learning \cite{lwf} has been a field of rising interest \cem{to address the aforementioned concerns.} 
The goal is to design learning algorithms that would continuously learn from a sequence of tasks to let the models imitate the learning process of human beings. A major problem that arises when a model is continually trained is called \emph{catastrophic forgetting} \cite{french}, which is associated with deteriorating performance on previously learned tasks.

Various solutions have been proposed to combat catastrophic forgetting in supervised learning settings,
including storage of previous data \cite{icarl, il2m, e2e-incremental, podnet}, or using generative models \cite{genreplay}, regularization of the model parameters between tasks \cite{ewc, gem, lwf, barber2018}, and the progressive growth of neural networks to combat forgetting \cite{mallya2017, fscil, prognets, expertgate}. These continual learning methods are applied in the image domain, however, continual learning has potential applications in speech and audio processing systems as well, including personal voice assistants, and conversational AI agents \cite{McTear2020ConversationalAD}. There indeed exists a few papers on supervised continual learning on audio, such as environmental sound classification \cite{gen_replay_audio, fewshot_audio_cl}, fake audio detection \cite{lwf_fake}, and audio captioning \cite{lwf_caption}. 

\cem{In this paper, we propose to adopt continual representation learning (CRL) in the sound event classification domain, and experimentally show its efficacy for class-incremental continual learning. Representation learning decouples the learning into two stages: a) Learning of an encoder with a representation-specific objective (e.g., with a self-supervised learning objective). b) Finetuning of a shallow classifier on top of the encoder for a downstream task. 
To the best of our knowledge, this is the first time that continual representation learning has been used in the sound event classification domain.
We argue that using a representation learning pipeline in continual learning is especially beneficial for practical use-case implications. Namely, i) the majority of the computational burden is passed on to the encoder learning stage, and therefore the shallow output layer can be trained including earlier tasks. ii) Decoupling the output head gives the additional flexibility to add an indefinite number of classes by re-training an ad-hoc output head for each task, which is required in real-life applications. iii) In the case where labeled data is scarce, learning an encoder from unlabeled data is beneficial for generalization performance, as we showcase with our experiments. }


As hinted above, we adopt self-supervised learning (SSL) within CRL. SSL is a relatively novel research area that is revolutionizing deep learning due to its impressive ability to learn features that generalize well without labels. In SSL, deep networks are trained with pretext tasks such as clustering \cite{deepclustering}, mutual information maximization \cite{dim}, image colorization \cite{colorization}, masked token prediction \cite{mae, bert}, contrastive learning \cite{moco,simclr, swav} and so on. These early works on self-supervised learning focused on images and natural language processing. Only recently has self-supervised learning been extended to audio and speech as well \cite{cpc, ssast, pase, wav2vec, byol_audio, bert_audio}.


In the space of incorporating representation learning within a continual learning setup, recent works include \cite{cassle, lump, lep, ssl_cil_omw}. Different from \cem{this paper}, these works are applied to images and involve explicit ways to combat catastrophic forgetting.
A continual representation learning method for speech recognition is proposed in \cite{cont_wav2vec} \zp{by learning more languages over time, but the size of the model grows as more languages are involved, and language information is required during inference.} 
\zp{In contrast, we focus on a more challenging setup where the model remains constant in size while progressively learning new sound classes, and no prior information on the input data is needed at test time (e.g., category of the sound).}

To summarize, in this paper we propose using the continual representation learning paradigm for class-incremental learning of new audio classes.
We list our contributions as follows:
\begin{itemize}
    \item \cem{We show that adopting representation learning for continual learning of new sound classes enables the study of the practically interesting case where only a small amount of labels is present. This is a realistic use case where the system is mainly trained on unlabeled data. 
    We investigate both in-domain and out-of-domain performance. For the evaluation of the partially labeled use case, we propose several evaluation metrics.} 
    

    \item \cem{We empirically observe that even if we do not employ an explicit mechanism to combat forgetting, employing similarity-based self-supervised learning within CRL yields better performance than continual supervised representation learning, and comparable performance to distillation-based continual learning methods to combat forgetting, which requires additional storage and computation.} 
    
    
\end{itemize}


 \vspace{-.35cm}
\section{Problem Statement}
\label{sec:prelim}

\subsection{Continual, Class-Incremental, Self-Supervised Learning}
\label{ssec:prelim_cl}
Continual learning (CL) aims to train a model on a sequence of datasets in a non-stationary training distribution. In this setup, the model is restricted from accessing any past or future data samples. Given an ordered sequence of tasks $\calT_1, \calT_2, \ldots, \calT_T$ with the associated datasets $\calD_1, \calD_2, \ldots, \calD_T$, the model is trained sequentially on one task at a time.    
In particular, we are interested in the class-incremental learning scenario \cite{classil} where any two tasks would not share a common class. 
During inference, the model is required to discriminate samples from all classes it has seen without knowing which task the test data comes from.


\cem{In this paper, we employ} similarity-based SSL methods \cite{moco, simclr, simclrv2, swav, barlowtwins}, \cem{where} we maximize the similarity between embeddings from a pair of distorted views of the same input to make representations robust to distortions.
We consider SimCLR \cite{simclr, simclrv2}, MoCo \cite{moco}, and Barlow Twins \cite{barlowtwins} as candidates for the continual representation learning framework. To create positive pairs for these similarity-based methods, we apply audio data augmentation by first taking fixed-length random segments from each clip, and then applying SpecAugment \cite{specaugment} to the spectrograms.





\subsection{Continual Representation Learning}
\label{ssec:prelim_crl}



\begin{figure*}[htb]
    \centering
    \includegraphics[width=0.8\textwidth]{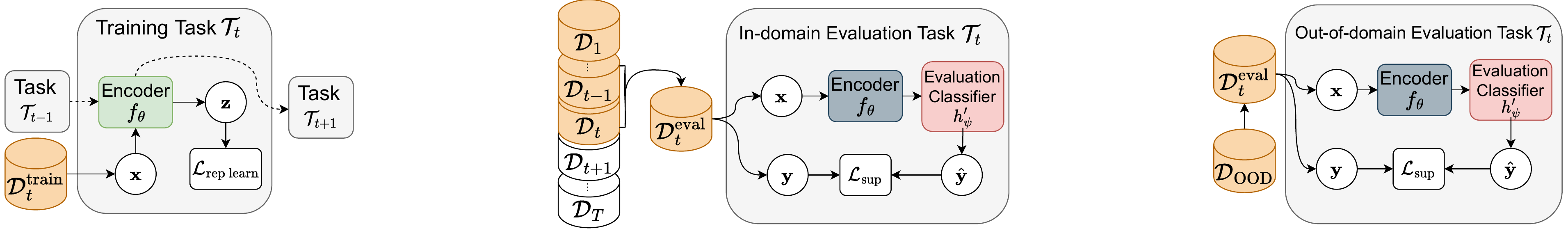}
    \caption{\textbf{(Left) Training pipeline for continual representation learning:} Representations are learned by continually training an encoder $f_\theta$ with one task/dataset at a time. \textbf{(Middle) In-domain evaluation:} The representations are evaluated by training a classifier $h_\psi'$ using a labeled training set on top of the frozen $f_\theta$. \textbf{(Right) Out-of-domain evaluation:} The dataset used for training the output head $h_\psi'$ is different from the dataset for encoder training.}
    \label{fig:pipeline}
\end{figure*}




State-of-the-art representation learning approaches require a large dataset to train high-quality representations. In practical cases where the dataset grows over time, frequently updating the model using all of the data samples is computationally prohibitive. 
The proposed continual representation learning framework circumvents this bottleneck by training an encoder only for the most recent task, which carries the majority of the computational burden. For instance, suppose we collect new data with new urban sound types (e.g., car horn, gunshot) to train a model that has been previously trained to recognize animal sounds (e.g., dog, rooster). Instead of retraining the network on both datasets, the encoder is updated only on the urban sound classes while preserving the ability to effectively represent animal classes. \cem{Once the encoder is updated with new \textbf{unlabeled} data, a shallow output layer is trained from scratch on top of the encoder, with available \textbf{labeled} data.}



That is, in continual representation learning, we aim to learn low-dimensional embeddings by training an encoder model $f_\theta$ on one task at a time. Under the class-incremental setting, in each task, the encoder is trained with data from a set of classes that corresponds to the current task: 
When entering a new task, $f_\theta$ is initialized from the end of the previous task. It is then updated with a representation learning objective with the current dataset $\calD_t^\text{train}$. 

After training, the learned representations are evaluated on a classification task. A shallow classifier $h_\psi'$ is trained on top of the encoder's output using a small amount of labeled data $\calD_t^\text{eval}$ with the encoder's weights fixed. In the case of in-domain evaluation, $\calD_t^\text{eval}$ contains examples from previous classes (see Figure \ref{fig:pipeline}), and in the case of out-of-distribution evaluation, $\calD_t^\text{eval}$ \zp{may come from a different distribution from the dataset used for training the encoder.}
This continual learning framework is shown in Fig. \ref{fig:pipeline}.


We propose using similarity-based self-supervised algorithms for representation learning in this framework, and we call this approach continual self-supervised representation learning (\textbf{CSSL}). 
We compare CSSL with continual supervised representation learning (\textbf{CSUP}), which relies on the assumption that annotations are accessible. In CSUP, we train an additional classifier $h_\psi$ jointly with $f_\theta$ to propagate label information. \zp{This classifier is discarded for downstream tasks or evaluation.} For a fair comparison, we apply identical data augmentations and encoder architecture to both approaches.

\zp{We hypothesize that similarity-based SSL is beneficial in the context of CRL due to its strong generalization ability.} 
The set of features learned from the current data generalizes well to past data, and consequently, this results in a system that is less prone to forgetting.

\vspace{-.1cm}
\begin{table}[!htb]
    \caption{Offline Classification Accuracy}
    \label{tab:offline}
    \centering
    \resizebox{0.9\linewidth}{!}{
    \begin{tabular}{c|c|c|c|c}
    & CSUP & SimCLR & Barlow Twins & MoCo \\
    \midrule
 UrbanSound8K & \textbf{80.9} & 74.3 & 73.3 & 69.3 \\
 DCASE TAU19 & \textbf{68.2} & 62.5 & 58.8 & 50.3 \\
\bottomrule
\multicolumn{5}{p{251pt}}{Offline accuracy when representations are trained with the entire dataset. }\\
    \end{tabular}
    }
\end{table}
\vspace{-.5cm}

\vspace{-.3cm}
\begin{table}[!htb]
\centering
\caption{In-domain Evaluation for CRL}
\resizebox{.5\textwidth}{!}{
\begin{tabular}{c|c|c|c|c||c|c|c|c}
\toprule
\multirow{3}{*}{Method} & \multicolumn{4}{c||}{UrbanSound8K ($T=5$)} & \multicolumn{4}{c}{DCASE TAU19 ($T=5$)} \\

& \multicolumn{2}{c}{LEP} & \multicolumn{2}{c}{SLEP} & \multicolumn{2}{c}{LEP} & \multicolumn{2}{c|}{SLEP}  \\
 & A ($\uparrow$) & F ($\downarrow$) & A ($\uparrow$) & F ($\downarrow$) & A ($\uparrow$) & F ($\downarrow$) & A ($\uparrow$) & F ($\downarrow$) \\
\midrule
\multicolumn{1}{c}{} & \multicolumn{8}{c}{\textbf{No distillation}} \\
CSUP & 65.6 & 19.6 & 48.4 & 33.1 & 48.2 & 27.6 & 32.5 & 38.4 \\
\rowcolor{Gray}
 SimCLR & \textbf{70.3} & 15.3 & \textbf{50.3} & 26.6& \textbf{59.7} & \textbf{17.8} & \textbf{42.1} & 27.9 \\
\rowcolor{Gray}
 Barlow Twins & 68.5 & \textbf{14.1} & 49.7 & \textbf{20.5} & 55.9 & 19.0 & 41.0 & \textbf{23.4} \\
 \rowcolor{Gray}
 MoCo & 68.4 & 15.6 & \textbf{50.3} & 25.8 &  49.5 & 20.2 & 34.8 & 25.8 \\
 \multicolumn{1}{c}{} & \multicolumn{8}{c}{\textbf{With distillation}} \\
 CSUP + $\calL_\text{MSE}$ & 58.6 & 27.0 & 43.8 & 39.2 & 49.1 & 26.1 & 35.1 & 35.8 \\
 CSUP + $\calL_\text{sim}$ & 70.6 & \textbf{13.8} & 54.9 & 27.1 &  56.2 & 19.7 & 42.1 & 32.4 \\
 CSUP + $\calL_\text{KLD}$ & 69.8 & 15.9 & \textbf{55.4} & 27.4 & 55.7 & 19.4 & 42.6 & 30.3 \\
 SimCLR + $\calL_\text{MSE}$ & \textbf{70.9} & 14.6 & 50.6 & 25.1 & 56.2 & 19.6  & 42.0 & 26.5 \\
 SimCLR + $\calL_\text{sim}$ & 70.6 & 14.0 & 51.1 & \textbf{25.0} & \textbf{60.0} & \textbf{17.6} & \textbf{42.8} & \textbf{25.9} \\
 \midrule

 \multicolumn{9}{p{301pt}}
 {Average accuracy (A) and forgetting (F) for CSSL and CSUP methods. Best performances are in bold. Note that rows with gray background correspond to the CSSL approaches where no-distillation against forgetting is used.}\\

\end{tabular}
}

\label{tab:cssl_csrl}
\end{table}

\section{Evaluation Protocol}
\label{sec:method_eval}

\vspace{-.05cm}
SSL algorithms are usually trained and evaluated in two steps \cite{simclr, moco, barlowtwins}. 
Similarly, in the continual learning setting, we first learn the representation with the encoder $f_\theta$ only using data from the current task. At the end of each task, we evaluate the representation by training an evaluation classifier, $h_\psi'$, randomly initialized, on top of the pre-trained frozen encoder (see Fig. \ref{fig:pipeline}). Note that $h_\psi'$ is much smaller in size compared to the encoder and is computationally more affordable to train. \zpCR{Also, notice that the classifier $h_\psi$ in CSUP (trained jointly with $f_\theta$) is discarded before evaluation since only the classes for the current task are seen during training while all previously seen classes are used for evaluation. Initializing a new evaluation classifier $h_\psi'$ from scratch ensures a fair comparison between CSSL and CSUP.}




Following supervised continual learning metrics for classification, for each task $t \in \{1, 2, \ldots, T\}$, we compute the accuracy of the classifier on the test set of task $j$ using the encoder at task $t$ denoted as $A_{t, j}$. We measure the average accuracy at the end of each task: $\bar{A_t} = \frac{1}{t} \sum_i A_{t,i}$ and obtain the average accuracy at the end of the training, $\bar{A} = \bar{A_T}$. Additionally, we compute forgetting, which measures the average decrease in accuracy of each task between its peak and the final performance, defined by $\bar{F} = \frac{1}{T - 1}\sum_{j=1}^{T-1} \max\limits_{\tau=1, 2, \ldots T} (A_{\tau, j} - A_{T, j})$.


    

We propose a set of evaluation protocols with different data distributions and types of models for the downstream classification task. \zp{When encoder training and evaluation take place on the same dataset (see Fig. \ref{fig:pipeline} Middle), we propose the following in-domain protocols: i) Linear Evaluation Protocol (\textbf{LEP}) \cite{lep}, where a linear classifier $h_\psi'$ is trained with the output of the fixed, pre-trained encoder $f_\theta$ using labeled data from past and current task $t$, $\{\calD_\tau \}_{\tau=1}^t$; ii) Subset Linear Evaluation Protocol (\textbf{SLEP}), where $h_\psi'$ is trained with a random subset, $\{\calD'_\tau\}_{\tau=1}^t$, where $\calD'_\tau \subset \calD_\tau$. This protocol simulates the scenario with limited labels. We keep a subset of 200 samples per task ($\approx 12\%$ of data). For both LEP and SLEP, classification becomes more challenging over time since the test data involves more classes as more tasks are learned.}

Additionally, we propose an out-of-domain (OOD) protocol (Fig. \ref{fig:pipeline} Right) where the representations are evaluated on a different dataset from which they are learned. We introduce the Full Linear Evaluation Protocol (\textbf{FLEP}), where at each task the linear evaluation classifier is trained on the full downstream dataset. As opposed to the in-domain protocols, here the test set does not grow with more tasks.


\vspace{-.4cm}
\section{Experimental Setup}
\label{sec:exp}


\vspace{-.3cm}
\subsection{Datasets}
\label{ssec:exp_ci}
\zp{We use UrbanSound8K \cite{us8k}, the TAU Urban Acoustic Scenes 2019 (DCASE TAU19) \cite{dcase_tau}, and VGGSound \cite{vggsound} to continually train the encoder.} We partition each dataset into class-disjoint subsets. Following \cite{classil}, we set the number of tasks to $T=5$ or $T=10$, and the classes are evenly split between tasks. Namely, each task corresponds to a set of classes. For example, in a $T=5$ case for a 10-class dataset, we have two distinct classes in each task. The ordering of classes is randomly shuffled before the split. \zp{The UrbanSound8K and the TAU Urban Acoustic Scenes 2019 (DCASE TAU19) are used for in-domain evaluation, and VGGSound \cite{vggsound} is used to learn representations for OOD experiments. We refer the readers to the appendix for more details on the datasets.}

\vspace{-.5cm}
\subsection{Model Training}
\label{ssec:exp_training}

We use the CNN14 architecture as the backbone encoder model with 78M trainable parameters \cite{pann}. The output representation has 2048 dimensions.
We train the encoder for 50 epochs per task on UrbanSound8K and VGGSound, and 100 epochs per task on DCASE TAU19. The linear classifiers for evaluation are trained for 30 epochs per task on UrbanSound8K and 100 epochs on DCASE TAU19.


Results are averaged across 10 folds on UrbanSound8K and across three runs with different seeds on DCASE TAU19 and VGGSound. Additional details can be found in our SpeechBrain \cite{speechbrain} implementation\footnote{\url{https://github.com/zhepeiw/cssl_sound}}.

\vspace{-.45cm}
\section{Results and Discussions}
\label{sec:res}

\vspace{-.2cm}
\subsection{Offline Evaluation}
\label{ssec:res_offline}
\zp{We first train and evaluate the representations with supervised and SSL algorithms under the ideal offline scenario in which the entire training set is available for training (i.e, there is only $T=1$ task). We consider three SSL algorithms, namely, SimCLR, MoCo, and Barlow Twins. For evaluation, we train a new classifier $h_\psi'$ on the output of the fixed, pre-trained encoder $f_\theta$ following Section \ref{sec:method_eval}.}

As we show in Table \ref{tab:offline} on both datasets, CSUP outperforms the CSSL approaches. This indicates that the supervised classification loss provides a strong baseline for representation learning when trained with the entire corpus at once without continual learning constraints.



\vspace{-.3cm}
\subsection{Comparison between CSSL and CSUP}
\label{ssec:res_cssl_vs_csup}

\zp{We compare representations learned from CSSL and CSUP in terms of their classification performance and resilience to forgetting. Similar to the offline experiments, we consider SimCLR, MoCo, and Barlow Twins for CSSL. In this experiment, we perform an in-domain evaluation with $T=5$ tasks, and we measure both values for average accuracy and forgetting at the completion of all tasks.}

\begin{figure}
    \centering
    \includegraphics[width=.21\textwidth]{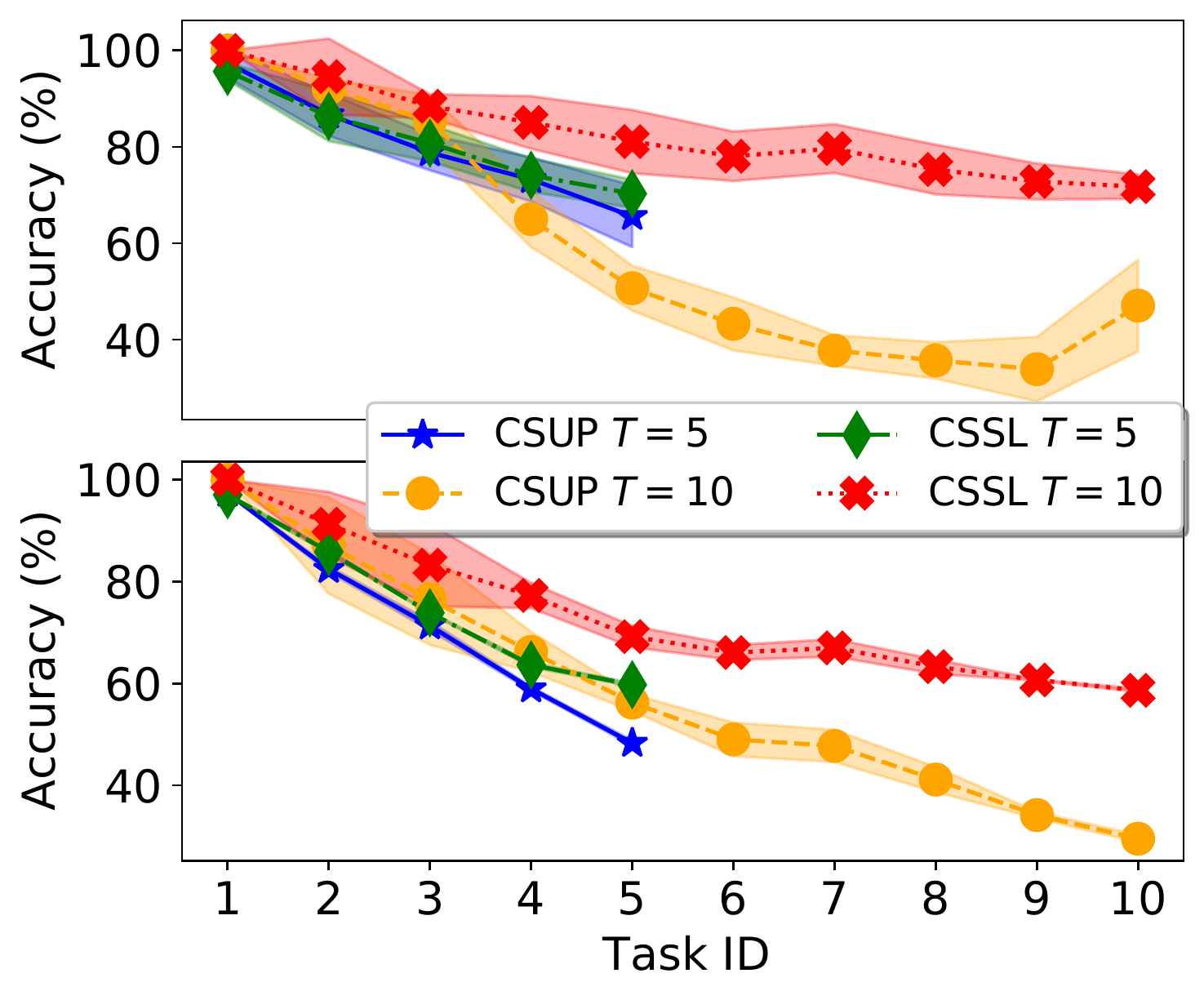}
    \includegraphics[width=.22\textwidth]{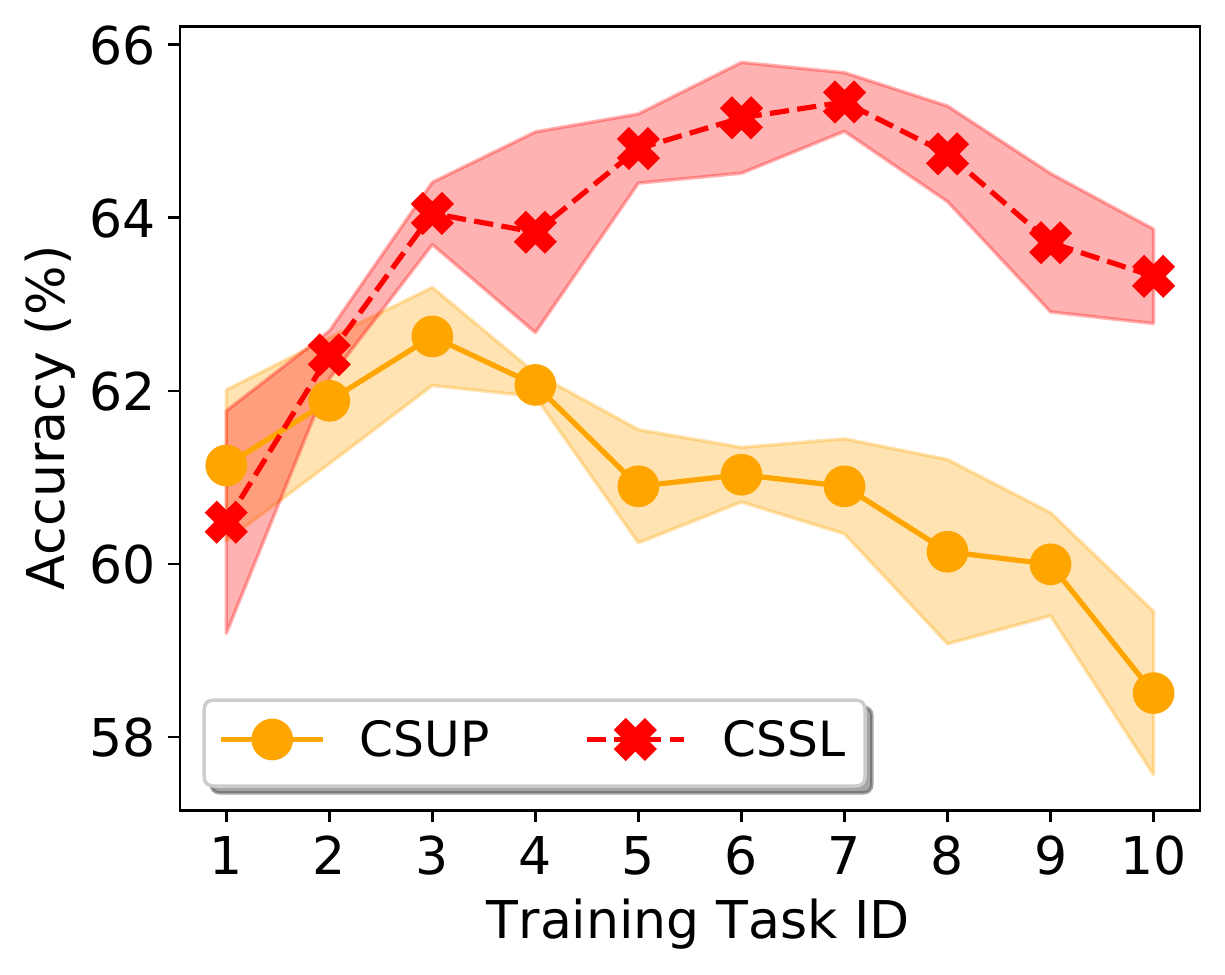}
    \vspace{-.2cm}
    \caption{\textbf{(Left) The In-Domain experiment:} The classification performance obtained with CSSL and CSUP on UrbanSound8K (top) and DCASE TAU19 (bottom). \textbf{ (Right) The OOD experiment:} Performance of CSSL and CSUP trained on VGGSound and evaluated with full-set linear evaluation protocol (FLEP) on DCASE TAU19.}
    \vspace{-.3cm}
    \label{fig:CL-results}
\end{figure}

Table \ref{tab:cssl_csrl} (No distillation) summarizes the average accuracy and forgetting for the in-domain protocols. First, the accuracy values for all learning methods for $T=5$ are lower than the corresponding offline accuracy in Section \ref{ssec:res_offline}; this indicates the existence of catastrophic forgetting for continual representation learning regardless of the approaches. While supervised learning beats SSL in offline evaluation, the trend is reversed when evaluating in a continual learning setup. All three CSSL methods (highlighted in gray) outperform CSUP on both datasets across all evaluation protocols and metrics. The consistent advantage of SSL in both accuracy and forgetting indicates that \textbf{the similarity-based objectives learn features that generalize better and are less affected by forgetting than CSUP}.

\zp{In Fig. \ref{fig:CL-results} (Left), we also show the trajectories of accuracies obtained with CSSL and CSUP for five tasks ($T=5$) and ten tasks ($T=10$) using SimCLR as the CSSL method. CSSL has higher average accuracy, and generally, the gap widens for both $T=5$ and $T=10$ as training progresses.}



\vspace{-.3cm}
\subsection{Knowledge Distillation}
\label{ssec:res_kd}

We additionally study the effects of 
regularization-based CL algorithms by adapting the Learning Without Forgetting (LwF) \cite{lwf} framework to the continual representation learning setting. At each task $t > 1$, \zp{the representation learning objective is optimized jointly with a knowledge distillation loss $\calL_\text{dist}(f_\theta(\bx), f_{\theta^{(t-1)}}(\bx))$, }
where $f_{\theta^{(t-1)}}$ is the snapshot of the encoder at the completion of task $t-1$ whose weights are frozen at the current task $t$. We consider the following candidates for the distillation loss $\calL_{\text{dist}}$: 
i) mean squared error, $\calL_\text{MSE}$ \cite{rh_free_dist}, ii) similarity-based objective that is used in SimCLR \cite{ntxent}, $\calL_\text{sim}$, and iii) KL-Divergence loss, $\calL_\text{KLD}$ (used for CSUP only) applied to the logits obtained after the output head \cite{lwf}. Table \ref{tab:cssl_csrl} (With distillation) shows the in-domain evaluation results of CSSL with SimCLR and CSUP with these loss options. \zp{To verify the consistency of these results for different SSL algorithms, we also tried Barlow Twins and MoCo and observed similar performance. We omit these results due to space constraints.}

We find that distillation with $\calL_\text{MSE}$ does not improve the results for CSUP on UrbanSound8K nor SimCLR on DCASE TAU19. $\calL_\text{KLD}$ outperforms $\calL_\text{MSE}$, but it is still beaten by the plain SimCLR except for the SLEP evaluation protocol, even though the plain SimCLR does not require storing any models or labels. Both $\calL_\text{MSE}$ and $\calL_\text{KLD}$ explicitly restrict the output to mitigate forgetting, but they reduce the plasticity of learning knowledge from new tasks and hence do not improve the overall performance. With the similarity-based loss $\calL_\text{sim}$, the accuracy of CSUP increases significantly on both datasets. The performance gain of CSUP once again highlights \zp{\textbf{the generalization ability of the similarity-based self-supervised objective and its resilience against forgetting in continual representation learning}}. The marginal improvement with $\calL_\text{sim}$ on SimCLR shows that the similarity-based framework alone learns features that generalize across tasks as well as distillation-based methods. For CSSL, distillation does not provide more advantages of generalization despite the cost of additional computation during training and storage for model saving. \textbf{Given the computational benefits, we hence conclude that CSSL without explicit methods for combating forgetting is preferable over the alternatives that use distillation. }



 \vspace{-.35cm}
\subsection{Out-of-domain (OOD) Evaluation}
\label{ssec:res_ood}

In addition to the in-domain evaluation, we compare CSSL and CSUP when the representation learning and downstream evaluation are performed on different datasets. Fig. \ref{fig:CL-results} (Right) shows the trajectories of the average accuracy using the FLEP protocol on DCASE TAU19 when the encoder is trained on VGGSound. 
We consider SimCLR for CSSL using $T=10$ tasks. When evaluating the representation using a linear classifier with the FLEP protocol, we see that CSSL outperforms CSUP after the second task. The performance gap widens as the encoder learns more tasks. We observe a decreasing trend in the accuracy curve of CSUP, indicating that the transfer of knowledge is overwhelmed by catastrophic forgetting. For CSSL, the curve keeps increasing for the first seven tasks. The results from the OOD evaluation
demonstrate the effectiveness of CSSL to learn continually without labels and its ability to generalize to a different downstream task. 

\vspace{-.55cm}
\section{Conclusion}
\label{sec:conc}
\vspace{-.2cm}

In this work, we propose a continual representation learning framework for sound classes. In this framework, the encoder training does not rely on labels and therefore is suitable for the practically relevant case where only a subset of labels is used to finetune an output classifier. Additionally, the framework is flexible enough to continually incorporate novel classes without the apriori knowledge of the total number of classes. With the continually pre-trained representations, the computational burden is significantly alleviated by simply finetuning a shallow classifier for downstream tasks. We showed, for the first time, that continual self-supervised learning gets competitive performance even if we do not use any mechanism against forgetting, which helps to reduce computational complexity. In future work, we plan to integrate continual self-supervised learning on more challenging audio tasks such as multi-label classification.

\bibliographystyle{IEEEtran}
\bibliography{refs}

\newpage
\section*{Appendix}
\label{sec:appendix}

\subsection{Dataset Details}
\label{ssec:appendix_dataset}

\subsubsection{UrbanSound8K}
\label{sssec:exp_dataset_us8k}


The UrbanSound8K dataset \cite{us8k} is a dataset for sound event recognition that contains 8,732 recordings with a total duration of 8.75 hours, where the length of each clip is limited to 4 seconds. Each recording is labeled with a single event class from the 10 possible classes,
where each class has no more than 1,000 clips. All files are pre-sorted into 10 folds for cross-validation.


\subsubsection{TAU Urban Acoustic Scenes 2019}
\label{sssec:exp_dataset_tau19}

The TAU Urban Acoustic Scenes 2019 dataset is used for the DCASE 2019 Task-1(A) challenge \cite{dcase_tau} for acoustic scene classification. The development set consists of 40 hours of audio collected from ten cities, with 9,185 recordings for training and 4,185 recordings for evaluation. Each clip is 10 seconds long and is labeled with one of 10 possible classes.

\subsubsection{VGGSound}
\label{sssec:exp_dataset_vggsound}

The VGGSound dataset \cite{vggsound} is an audio-visual dataset with 200,000 clips with a total duration of 560 hours from more than 300 sound classes such as instruments, horns, and city sounds.
The recordings are scraped from YouTube and are labeled by pre-trained image and sound classifiers. However, these data samples may not be accurately labeled due to i) the classifiers are pre-trained on a different dataset, and ii) the recordings may contain interfering sound events while each clip is assigned a single label. We consider learning the representation using VGGSound while evaluating with a different downstream dataset with an OOD setup.


\vspace{-.3cm}

\subsection{Continual Representation Learning with Full Replay Buffer}
\label{ssec:csup_all_data}

\zpCR{We also consider continually training the encoder $f_\theta$ and the output head $h_\psi$ with a full replay (FR) buffer by storing data samples from the current and all previous tasks with $T=5$ tasks. We provide the final average accuracy using the in-domain linear evaluation protocol for UrbanSound8K and DCASE TAU19 in Table \ref{tab:full_buffer} for CSUP and CSSL using SimCLR, denoted by CSUP-FR and SimCLR-FR, respectively. We also include the corresponding offline systems (denoted by CSUP-O, SimCLR-O) from Table \ref{tab:offline} and the continually trained systems without using replay buffer (trained only on the current task) - (denoted by CSUP-NR, SimCLR-NR) taken from Table \ref{tab:cssl_csrl}. Both FR systems achieve the best performance compared to the offline and NR systems by beating the offline systems by a small margin (possibly due to a curriculum effect) and significantly exceeding the accuracy of the NR systems. When the encoder observes data from all tasks in both offline and full replay scenarios, the supervised algorithm slightly outperforms SimCLR. However, this performance gain comes at the extra computational cost of storing and revisiting past data samples during the encoder training. Also notice that the gap between CSUP-FR and CSUP-NR is significantly larger than the one between SimCLR-FR and SimCLR-NR, and this further indicates that representations learned with supervised objectives are more prone to performance degradation when access to previous data is restricted.}

\begin{table}[!htb]
    \caption{In-domain Evaluation for Different Encoder Training Dataset}
    \label{tab:full_buffer}
    \centering
    \resizebox{1.0\linewidth}{!}{
    \begin{tabular}{c|c|c|c|c|c|c}
   & CSUP-O & SimCLR-O & CSUP-NR & SimCLR-NR & CSUP-FR & SimCLR-FR \\
    \midrule
   US8K & 80.9 & 74.3 & 65.6 & 70.3 & \textbf{82.3} & 77.2 \\
   DCASE & 68.2 & 62.5 & 48.2 & 59.7  & \textbf{69.1} & 67.4 \\
\bottomrule
\multicolumn{7}{p{321pt}}{Final average accuracy with the linear evaluation protocol when representations are trained with the full dataset at once (CSUP-O, SimCLR-O), continually trained without using data buffer (CSUP-NR, SimCLR-NR), and using full data buffer (CSUP-FR, SimCLR-FR) for $T=5$ tasks.}\\
    \end{tabular}
    }
\end{table}
\vspace{-.5cm}

\begin{figure}[htb]

\begin{minipage}[b]{1.0\linewidth}
  \centering
  \centerline{\includegraphics[width=1.0\textwidth]{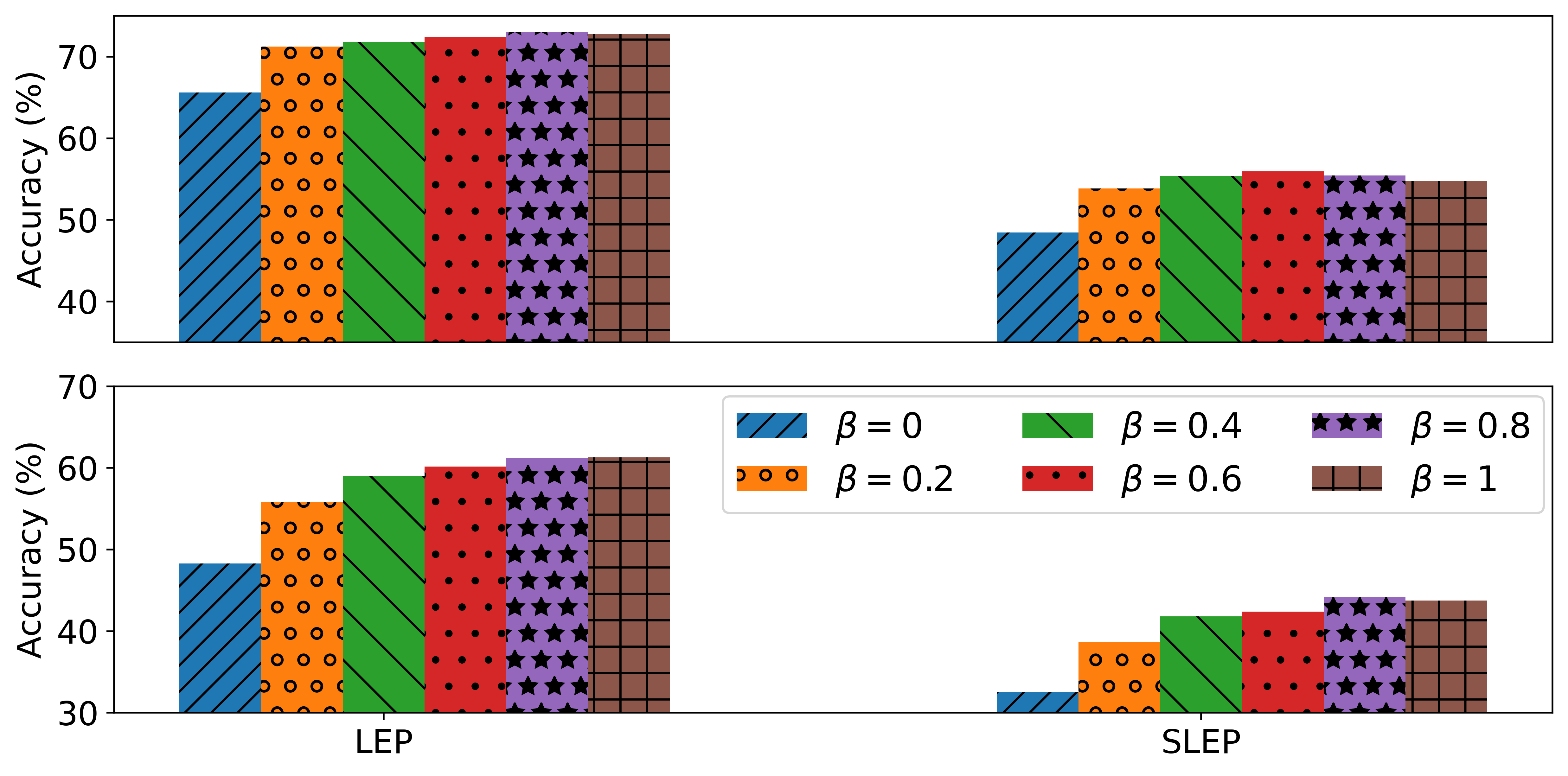}}
\end{minipage}
\vspace{-.9cm}
\caption{Average accuracy on the linear combination of CSSL and CSUP with the supervised weight $\alpha=1$ fixed on in-domain linear evaluation (LEP) and subset evaluation (SLEP) protocols on UrbanSound8K (upper) and DCASE TAU19 (lower).}
\label{fig:ablation}
\end{figure}

\subsection{Assessing the Impact of Self-Supervised Objectives}
\label{ssec:appendix_joint_ssl}

To analyze the influence of the SSL objective on continual representation learning, we first perform an ablation study by controlling the amount of self-supervision in the learning framework. For this experiment, we assume the existence of annotations and train the encoder jointly with both classification loss and the similarity-based SSL loss \cite{ntxent}, given by $\calL_\text{joint} = \alpha\calL_{\text{sup}} + \beta\calL_{\text{ssl}},$
where $\alpha, \beta \geq 0$ control the weight of each objective. We fix the weight of the supervised loss as $\alpha=1$ and increase the weight of the SSL objective $\beta$ from 0 to 1. Notice that $\alpha=1, \beta=0$ is equivalent to CSUP. We compute the final average accuracy at the end of $T=5$ tasks using the in-domain protocols on two different datasets, as shown in Figure \ref{fig:ablation}. In general, higher weights on SSL leads to better performance. In particular, there is a significant performance gain between $\beta=0$ and $\beta=0.2$ across both protocols. These observations imply the benefits of incorporating similarity-based SSL objectives in continual representation learning even if label information is available.

\subsection{Impact from the Number of Tasks}
\label{ssec:appendix_numtask}

We are interested in whether the relative performance between CSSL and CSUP is consistent with a varying total number of tasks. As mentioned in Section \ref{ssec:res_cssl_vs_csup}, Fig. \ref{fig:CL-results} (Right) displays the trajectories of the average accuracy at the end of each task for CSSL using SimCLR and CSUP with $T=5, 10$ total tasks. CSSL has a higher average accuracy at the completion for both $T=5, 10$ on both datasets than CSUP. The performance gap is significantly larger in $T=10$ than in $T=5$. Furthermore, in $T = 10$, this gap widens as training progresses. Since both UrbanSound8K and DCASE TAU19 contains 10 classes, $T=10$ creates an extreme setting where the encoder learns only one class at a time. If training without replay data using cross-entropy, the supervised framework may optimize by biasing the weight towards the current class without learning useful representations. In contrast, without label information, the similarity-based CSSL framework is more robust to the shift in data distribution across tasks.

\end{document}